\documentclass[AMA,Times1COL]{WileyNJDv5}

\articletype{Article Type}%
\received{00} 
\revised{00} 
\accepted{00}

\usepackage{moreverb,url}
\usepackage{amsmath}
\usepackage{mathtools}
\usepackage{bm}
\usepackage{multirow}
\usepackage{pdflscape}
\usepackage{booktabs}
\usepackage{tablefootnote}
\usepackage{threeparttable}
\usepackage{subfig} 
\usepackage[english]{babel}
\usepackage{amsthm}

\usepackage[figuresright]{rotating}

\let\code=\texttt

\newcommand{\E}{\mathrm{E}}

\newcommand{\HR}{\mathit{HR}}
\newcommand{\WP}{\mathit{WP}}
\newcommand{\RWP}{\mathit{RWP}}

\newcommand{\FWR}{\mathit{FWR}}

\newcommand{\MST}{\mathit{MST}}
\newcommand{\RMST}{\mathit{RMST}}

\newcommand\BibTeX{{\rmfamily B\kern-.05em \textsc{i\kern-.025em b}\kern-.08em
T\kern-.1667em\lower.7ex\hbox{E}\kern-.125emX}}

\begin{document}

\title{Restricted Win Probability with Bayesian Estimation for Implementing the Estimand Framework in Clinical Trials With a Time-to-Event Outcome}

\author[1]{Michelle Leeberg} 
\author[1,2]{Xianghua Luo}
\author[1]{Thomas A. Murray} 

\authormark{Leeberg et al.} 

\address[1]{\orgdiv{Division of Biostatistics and Health Data Science}, \orgname{School of Public Health, University of Minnesota}, \orgaddress{\state{MN}, \country{USA}}}
\address[2]{\orgdiv{Biostatistics Core, Masonic Cancer Center}, \orgname{University of Minnesota}, \orgaddress{\state{MN}, \country{USA}}}


\corres{Thomas Murray, murra484@umn.edu}


\abstract[Abstract]{We propose a restricted win probability estimand for comparing treatments in a randomized trial with a time-to-event outcome. We also propose Bayesian estimators for this summary measure as well as the unrestricted win probability. Bayesian estimation is scalable and facilitates seamless handling of censoring mechanisms as compared to related non-parametric pairwise approaches like win ratios. Unlike the log-rank test, these measures effectuate the estimand framework as they reflect a clearly defined population quantity related to the probability of a later event time with the potential restriction that event times exceeding a pre-specified time are deemed equivalent. We compare efficacy with established methods using computer simulation and apply the proposed approach to 304 reconstructed datasets from oncology trials. We show that the proposed approach has more power than the log-rank test in early treatment difference scenarios, and at least as much power as the win ratio in all scenarios considered. We also find that the proposed approach's statistical significance is concordant with the log-rank test for the vast majority of the oncology datasets examined. The proposed approach offers an interpretable, efficient alternative for trials with time-to-event outcomes that aligns with the estimand framework.}


\footnotetext{Bayesian Restricted Win Probability}
\maketitle

\section{Introduction}\label{section intro}
In phase III randomized controlled trials (RCTs), time-to-event endpoints are often used to assess the efficacy of the active treatment group(s) compared to the control group. Overall survival is the preferable primary endpoint in oncology \cite{fda_guidance_2007}. Other common time-to-event endpoints in RCTs include progression-free survival, disease-free survival, and event-free survival \citep{delgado_clinical_2021}. A recent search of ClinicalTrials.gov identified 756 phase III trials between 2017 and 2022 that used a survival endpoint \cite{noauthor_search_nodate-1}. 

 The log-rank test facilitates evaluation of treatment efficacy with respect to a time-to-event endpoint \citep{mantel_statistical_1959}. This very common approach is nonparametric and most powerful under proportional hazards (PH) \citep{karadeniz_examining_2017}. To communicate the treatment effect, the log-rank test may be supplemented with a hazard ratio (HR) estimate from a Cox PH model as the log-rank test is equivalent to the score test from a Cox PH model \citep{collett_modelling_2023}. An addendum to the Steering Committee of the International Conference on Harmonization (ICH) E9 guidelines \cite{ich_e9_working_group_ich_2017} was released to the public in 2017 discussing estimands in the context of RCTs. The estimand framework requires a clearly defined, clinically meaningful summary measure for comparing treatments. 
 Rufibach \cite{rufibach_treatment_2019} discusses the estimand framework in regards to time-to-event outcomes and notes that the log-rank test and corresponding HR depend on the censoring distribution when PH fails to hold. The hazard ratio will also depend on the follow-up time and may not have a causal interpretation even in RCTs as period specific hazard ratios exhibit selection bias due to only individuals who have not had an event up to that time point included \citep{hernan_hazards_2010, aalen_does_2015}. HRs also are commonly misinterpreted, especially when PH does not hold. In a 2019 study by Weir et al., 47.2\% of the medical residents, fellows, and corresponding authors for randomized trials misinterpreted HRs \citep{weir_interpretation_2019}.  
 
 Some alternative approaches that better align with the estimand framework use summary measures such as survival at a milestone time (e.g., 5-year survival), median survival, or restricted mean survival time (RMST) for treatment comparisons. Milestone survival, while easy to interpret, does not account for the timing of events and lacks power. Median survival is only identifiable when the cumulative incidence is at least 50\%, and it also lacks power. Difference in RMST offers a more efficient summary measure that aligns with the estimand framework and does not rely on the PH assumption \citep{royston_restricted_2013, trinquart_comparison_2016, freidlin_methods_2019}. 
 Its interpretation as the expectation of the event time or $\tau$, whichever is smaller, is also challenging. Windows mean survival time is an extension of RMST, developed for late effects cases where RMST has low power, that includes a lower limit, which further complicates interpretation \citep{paukner_versatile_2022}. 

There remains a need for efficient, clinically meaningful summary measures that facilitate applying the estimand framework in clinical trials with time-to-event endpoints but do not require strong parametric assumptions about the treatment effect. In this paper, toward this aim, we propose and evaluate restricted and unrestricted \textit{win probability} (WP) estimands. The WP summary measure reflects the probability that an individual assigned to an active treatment will experience a later event time compared to an independent individual assigned to the control, with an optional restriction that all events occurring after a specified milestone time point, e.g., the trial follow-up period, are considered a tie. The unrestricted WP estimand was introduced by Acion et al. as the probabilistic index, however, the estimand was discussed in the context of continuous outcomes \cite{acion_probabilistic_2006}. To facilitate estimation and inference, we will evaluate the proposed measures using a Bayesian model that does not assume PH. Since these are distributional summary measures, they may be evaluated using any generative model framework, including parametric or non-parametric approaches. Using a Bayesian approach, the proposed WP summary measures will inherit a posterior distribution with inference following in the usual manner.

Our proposed WP summary measures with Bayesian estimation are related to, but distinct from the frequentist \textit{win ratio} (FWR) and generalized pairwise comparisons approaches, which use pairwise comparisons of the observed outcomes between two groups to tabulate the numbers of winning, losing, and tied pairs and thereby assess efficacy \citep{buyse_generalized_2010, pocock_win_2012, zheng_win_2023}. Other generalized pairwise comparison approaches include net benefit, win odds, probabilistic index, Finkelstein–Schoenfeld method, and O'Brien's test \citep{deltuvaite-thomas_generalized_2023, brunner_win_2021}. Critically, no generalized pairwise comparison methods include the restricted win probability measure as an estimand \cite{deltuvaite-thomas_generalized_2023, ozenne_asymptotic_2021}. These win probability estimands also have not been estimated using a Bayesian approach, which facilitates seemless handling of censoring mechanisms and incorporation of prior information. In particular, we fit a flexible Bayesian model for the two survival distributions that does not assume proportional hazards, and then apply the relevant posterior sample transformations to carry out estimation and inference on the restricted and unrestricted win probability estimands. In this way, the Bayesian estimators scale linearly with sample size rather than exponentially as with the frequentist non-parametric pairwise estimators. 

In this paper, we will compare the proposed Bayesian WP approaches with the log-rank test, RMST, and FWR methods using computer simulations. We will further compare the proposed approaches with the log-rank test and FWR method by applying them to datasets from 153 phase III oncology trial publications, representing 142 distinct trials, available in the \texttt{KMdata} R package, which were reconstructed from Kaplan-Meier curves via the Guyot algorithm \citep{guyot_enhanced_2012}. 

The rest of this paper is organized as follows. In Section~\ref{sectie RI model},
we define the proposed WP summary measures, discuss established measures, and introduce the Bayesian model we will use to evaluate the proposed measures. In Section~\ref{sims},
we describe the design of the simulation study and present simulation results. In Section~\ref{real dat},
we analyze data from 153 phase III oncology clinical trial publications. We finish by discussing practical implications of this research in Section~\ref{discussion}.

\section{Methods}\label{sectie RI model}
We first introduce a general definition of the win probability, and then define the proposed WP estimand for a univariate time-to-event outcome. We then discuss our Bayesian approach for estimation and the specific probability models we use in this paper. 

\subsection{Win probability estimand}

The win probability estimand comparing two treatments in an RCT
reflects the probability that a participant assigned to an active treatment will have a better outcome than an independent participant assigned to the control treatment with ties broken evenly.
This estimand is also referred to as the Mann-Whitney estimand or probabilistic index \citep{acion_probabilistic_2006, fay_causal_2018}. 
Let $Y_{a}$ and $Y_{c}$ denote independent random variables with support $\mathcal{Y}$ following respective distributions under the active ($a$) and control ($c$) treatment conditions. When ties are not possible, the active treatment will \textit{win} whenever $Y_{a}$ takes a value that is preferable to that of $Y_{c}$, and otherwise the active treatment will incur a \textit{loss}. The WP estimand comparing $a$ to $c$ arises as $\WP(a,c)=\Pr(Y_{a} \text{ wins versus } Y_{c})$ such that $\WP(a,c) = 1 - \WP(c,a)$ with a null value of 0.5. 
A \textit{tie} would be assigned when $Y_a = Y_c$, or in certain cases when $Y_a$ and $Y_c$ are not equal but clinically equivalent as determined in the specific context. While the WP with ties no longer has a simple probabilistic interpretation, we extend its definition in the usual way as $\WP(a,c)=\Pr(Y_{a} \text{ wins versus } Y_{c}) + 0.5 \Pr(Y_{a} \text{ ties } Y_{c})$ such that $\WP(a,c) = 1 - \WP(c,a)$ with a null value of 0.5 as before with no ties.
In general, $\WP(a,c)$ approaching 1 indicates that treatment $a$ is superior to treatment $c$, as it is increasingly likely that a patient's outcome under treatment $a$ will be preferable to another patient's outcome under treatment $c$. Similarly, $\WP(a,c)$ close to 0 indicates that treatment $a$ is inferior to treatment $c$.
Summary measures that one could report instead of or in addition to the proposed win probability include \textit{net benefit}, $\WP(a,c) - \WP(c,a)$, and \textit{win odds}, $\WP(a,c)/\WP(c,a)$ \citep{buyse_generalized_2010,brunner_win_2021}.

One can define $\WP(a,c)$ with respect to a pairwise \textit{win function} $r:\mathcal{Y} \times \mathcal{Y} \to \{1,0.5,0\}$ where $Y_k \in \mathcal{Y}$, $k=a,c$,  and a win, tie, and loss are quantified as 1, 0.5, and 0, respectively.  Given $r$, the WP summary measure arises as 
\begin{equation}
\WP(a,c)=\E[r(Y_{a},Y_{c})] =\Pr(Y_{a} \text{ wins versus } Y_{c}) +  0.5 \Pr(Y_{a} \text{ ties } Y_{c}).
\end{equation}
In this way, the proposed WP metric builds upon \textit{Bayesian utility-based designs}, which have been used to determine treatment efficacy by eliciting a clinical utility function that quantifies the relative desirability of every outcome value and comparing groups with respect to expected utility \citep{murray_utility-based_2016, murray_robust_2017, murray_utility-based_2018}. In fact, RMST is a special case of a utility-based summary measure for time-to-event outcomes. Utility-based comparisons are limited by their intrinsic subjectivity, wherein inference may be sensitive to modest changes in the specified clinical utilities. The proposed WP approach instead relies on a win function that may be easier to specify than a clinical utility function as the win function only requires a partial ordering of $\mathcal{Y}$ rather than a full quantitative clinical valuation of $\mathcal{Y}$, which reduces the subjectivity involved in the treatment comparison and thereby improves generalizability. 


For the remainder of the paper, we will apply the WP estimand in the context where $Y_{k}$ is a continuous time-to-event random variable subject to right censoring with $\mathcal{Y} \equiv \mathcal{R}^+$. The unrestricted WP estimand is defined with respect to
\begin{equation}
   r(Y_{a},Y_{c}) = \left\{
     \begin{array}{lr}
       1 & \text{if } Y_{a}>Y_{c}, \\ 
      0 & \text{if } Y_{a}<Y_{c}.
     \end{array}
     \right.
\end{equation}

\noindent Because $Y_{k}$ is continuous, ties are not possible, i.e., $\Pr(Y_{a}=Y_{c}) = 0$, and the corresponding unrestricted WP estimand arises as, 
 \begin{equation}
     \WP(a,c) = \E[r(Y_{a},Y_{c})]=\Pr(Y_{a}>Y_{c})=\int_{0}^{\infty}S_{a}(t)f_{c}(t)dt,\label{WP}
 \end{equation}
where $f_{k}(t)$ is the probability density function, $S_{k}(t) = 1-F_k(t)$ is the survival function, and $F_k(t)$ is the distribution function for $Y_k$, $k=a,c$. 

In clinical trials with a primary time-to-event endpoint, every participant may be followed for a specific period of time, e.g., 90 days, or for a minimum period of time, e.g., 2 years, allowing longer follow-up for participants who enrolled early.
The specific follow-up duration usually depends on the condition and outcome of interest, the trial population, and funding, among other factors. The unrestricted $\WP(a,c)$ estimand reflects the full survival distribution, which in most cases will not be discernible without extrapolation beyond the maximum follow-up period, and thus the estimate of the unrestricted $\WP(a,c)$ may be sensitive to the estimates of the tails of $F_{a}$ and $F_{c}$.
To preclude the need for extrapolation and improve robustness, we propose the restricted win probability (RWP) defined with respect to
\begin{equation}
   r_{\tau}(Y_{a},Y_{c}) = \left\{
     \begin{array}{ll}
       1 & \text{if } Y_{a}>Y_{c} \text{ and } Y_{c} \leq \tau, \\
        0.5 & \text{if }  Y_{a} > \tau  \text{ and } Y_{c} > \tau,\\
        0 & \text{if } Y_{a}<Y_{c} \text{ and } Y_{a} \leq \tau,
     \end{array}\label{Function_r2}
   \right.
\end{equation} 
where $\tau$ is prespecified, say, as the minimum administrative follow-up duration. 
That is, $r_\tau$ declares a tie when $Y_a>\tau$ and $Y_c>\tau$, and declares a winner only when either $Y_a$ or $Y_c$ is less than $\tau$. 
The corresponding RWP estimand arises as follows (see Web Appendix A for the detailed derivation), 
\begin{equation}
\begin{split}
    \WP_{\tau}(a,c) =\E[r_{\tau}(Y_{a},Y_{c})]
    &=\Pr(Y_{a}>Y_{c} \text{ and } Y_{c} < \tau)+0.5\Pr(\min(Y_{a}, Y_{c})>\tau)\\
    &=\int_{0}^{\tau}S_{a}(t)f_{c}(t)dt + 0.5S_{a}(\tau)S_{c}(\tau).\label{RWP}
\end{split}
\end{equation}




Similar to difference in RMST between treatments, the proposed estimands in Equations (\ref{WP}) and (\ref{RWP}) are widely applicable as they do not require restrictive assumptions about the form of the treatment effect (e.g., PHs) and incorporate more information about the distributions than a milestone time or median comparison. In contrast to the HR from a Cox PH model and FWR, the proposed estimands do not depend on the censoring distribution.


\subsection{Bayesian modeling and estimation} 
The proposed win probability estimands are statistical functionals of the unknown distribution functions $F_a$ and $F_c$. 
In contrast with the FWR, which relies on pairwise comparisons of the observed outcomes for estimation, we will use a Bayesian approach for estimation by specifying a probability model for $Y_{a}$ and $Y_{c}$, indexed by a vector of parameters denoted by $\bm{\theta}$. 
The Bayesian approach allows seamless handling of censoring mechanisms via established methods for Bayesian analysis and imposes a specific definition for $\WP(a,c)$ under the assumed model. To distinguish between the population estimands defined in Equations (\ref{WP}) and (\ref{RWP}), and their Bayesian estimators, we denote the resulting Bayesian model-based estimators as $\WP(a,c;{\bm{\theta}})$ and $\WP_\tau(a,c;{\bm{\theta}})$. We will use the posterior probability $\Pr(\left.\WP(a,c;{\bm{\theta}}) > 0.5 \right|\textrm{Data})$ to quantify the evidence for whether treatment $a$ is superior to treatment $c$. 

For the remainder of this paper, following Ibrahim et al \citep{ibrahim_bayesian_2010}, we assume 
\begin{eqnarray*}
        Y_{k}|\nu_{k},\lambda_{k}&\sim& \textrm{Weibull}(\nu_{k},\lambda_{k}),\\
    \nu_{k}&\sim& \textrm{Gamma}(0, 0.0001),\\
    \lambda_{k}&\sim&  \textrm{Log Normal}(1.0005, 0.0001),
\end{eqnarray*}
with $f(y|\nu,\lambda)=\nu\lambda y^{\nu-1}\exp(-\lambda y^{\nu})$ and $S(y|\nu,\lambda)=\exp(-\lambda y^{\nu})$. We allow distinct shape and scale parameters for $Y_a$ and $Y_c$, i.e., $(\nu_a,\lambda_a)\perp(\nu_c,\lambda_c)$, such that our assumed model affords more flexibility with respect to the treatment effect than parametric PH or accelerated failure time (AFT) models. 
One may express and evaluate $\WP(a,c;\bm{\theta})$ and $\WP_{\tau}(a,c;\bm{\theta})$ using any Bayesian model suitable for time-to-event random variables $Y_a$ and $Y_c$, including piecewise exponential models and non-parametric approaches \citep{ibrahim_bayesian_2010, muliere_bayesian_1997, damien_bayesian_2002}. However, a Weibull distribution can exhibit a variety of shapes \citep{guure_bayesian_2012, guure_bayesian_2013} and facilitates posterior computation of the proposed estimators. Posterior samples of $\bm{\theta}$ are transformed into posterior samples of the win probability estimands via numerical integration with adaptive quadrature (using \code{integrate} function in R package \texttt{stats}). 
All analyses were done using R version 4.0.4 and JAGS version 4.3.0.  

\section{Simulation Studies }\label{sims}

\subsection{Simulation study design \label{design}}
Our simulation study evaluates 
the restricted or unrestricted WP as the primary comparative summary measure for a two-arm RCT as compared with the log-rank test, RMST, and FWR. For the restricted WP, we set $\tau$ to the minimum administrative follow-up time. We implemented RMST using the \texttt{rmst2} function in the \texttt{survRM2} R package and set $\tau$ to the smaller of the minimum administrative follow-up time and the last event time in each randomization group. We implemented FWR using the \texttt{WRrec} function in the \texttt{WR} R package with the last-event assisted win ratio being reported. 






We compared all of the approaches with respect to power and two-sided type I error rate. 
However, because each approach targets a different estimand, we compared the performance of only the WP estimators with respect to relative bias, root mean squared error (RMSE), 95\% credible interval (CI) coverage, and 95\% CI width. 
We estimated these performance metrics using Monte Carlo simulation methods. 

\begin{figure}
    \centering
    \includegraphics[width=0.8\textwidth]{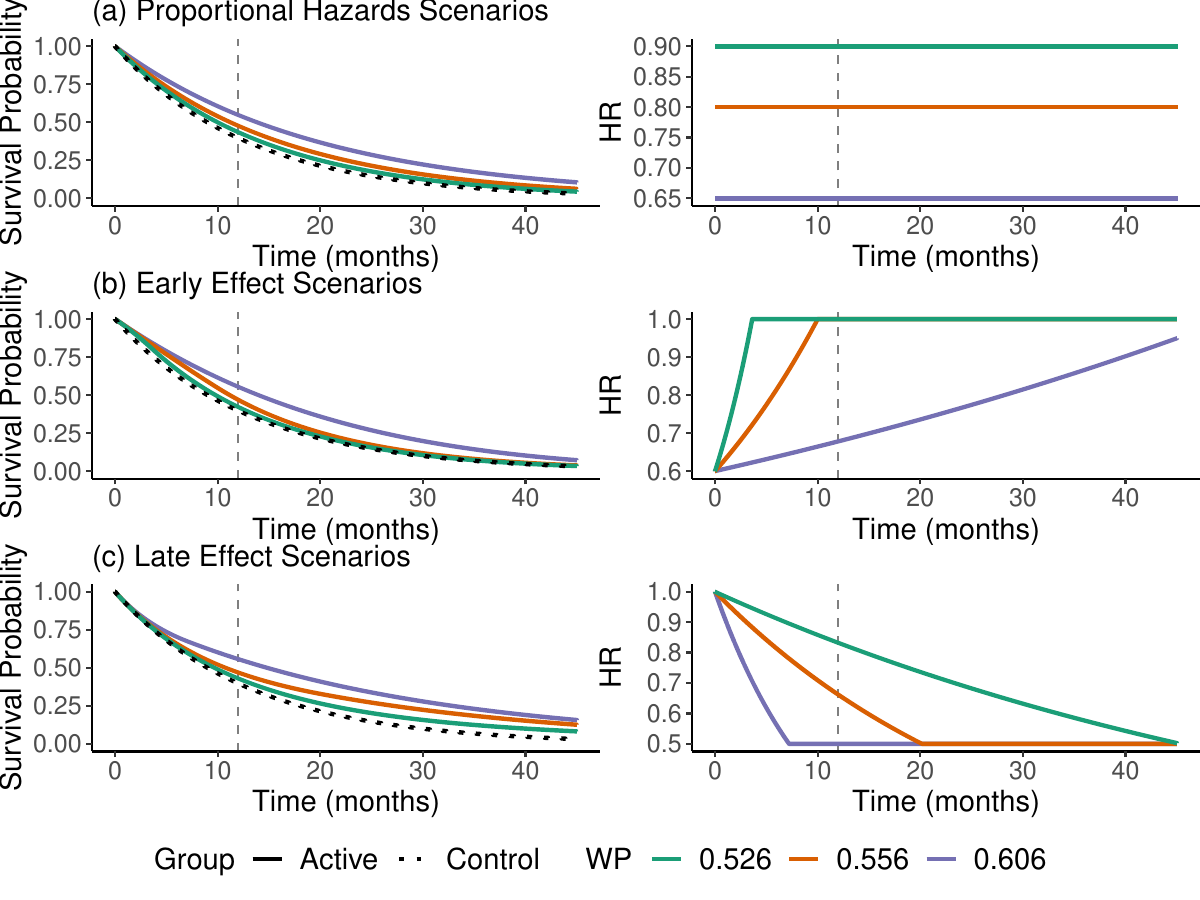}
    \caption{Survival probability curves and hazard ratio (HR) plots for all three simulation scenarios. The rows are for each scenario, with the left column for survival curves and the right column for HRs. In all scenarios, the control group follows an exponential distribution with a median survival time of 9 months. In row (a), treatment groups follow exponential distributions with proportional hazards to the control group. In rows (b) and (c), treatment groups have log-linear hazard functions; with (b) having early differences in hazard and (c) having late differences in hazard. Within each plot, three levels of treatment effect as measured by the win probability (WP) are presented, while the vertical dashed line indicates the follow-up time of 12 months, $\tau=12$ for studying the restricted WP.}
    \label{surv_hr_fig}
\end{figure}

We used 1:1 treatment allocation throughout. We considered three outcome data generative scenarios where the control arm outcomes always follows an exponential distribution with a median survival time of 9 months and the treatment arm outcome distribution is chosen to exhibit different types of treatment effects, including PH, non-PH exhibiting an early effect, and non-PH exhibiting a late effect. We created the non-PH effects by specifying a piecewise linear log-hazard function in the treatment arm (see Web Appendix B for simulation setting details).
For each type of treatment effect, we manipulated the data generative parameters to exhibit three effect sizes with corresponding $\WP$ values of 0.606, 0.556, and 0.526. Figure \ref{surv_hr_fig} depicts the assumed true survival curves and HRs for the resulting scenarios (see Web Figure 1 for the hazard function curves).

Table \ref{est_rel_table} depicts the magnitude and relationship of various treatment effect measures under the true model: $\WP(a,c)$ and $\WP_{\tau}(a,c)$, accompanied by $\mathit{HR}$ (or average HR up to $\tau$ when the PH assumption does not hold), and difference in RMST, median and mean survival time (denoted by $\Delta \RMST$, $\Delta \MST$ and $\Delta\mu$, respectively), for the different simulation scenarios. In the null scenario, $\WP(a,c) = \WP_{\tau}(a,c) = 0.5$, while in the other scenarios $\WP(a,c)>\WP_{\tau}(a,c)$ with larger values also corresponding to larger RMST differences, and differences in median or mean survival time. We include an average HR for non-proportional hazards scenarios, however, this may not be equal to the HR estimated by a Cox PH model which is dependent on the censoring distribution.

We selected the largest simulated sample size in each scenario to ensure the log-rank test would provide approximately 80\% power for the smallest effect size.
To mimic a setting where only a moderate proportion of the survival distribution is identifiable, all individuals were uniformly censored between 12 and 21 months. We present additional simulation results where a larger proportion of the survival distribution is identifiable in Web Appendix B,
namely where individuals were uniformly censored between 36 and 45 months. For each scenario and effect size, we generated 2000 datasets. For each dataset, we carried out inference for the proposed summary measures using 5000 posterior draws. Further technical details about the data generative assumptions for each simulation scenario are provided in Web Appendix B.

\begin{table}[ht]
\centering
    \caption{For the different simulation scenarios and a null scenario, unrestricted $\WP(a,c)$ and restricted win probability $\WP_{\tau}(a,c)$ with $\tau=12$, the active ($a$) and control ($c$) groups' hazard ratio ($\HR$; for Early and Late scenarios, it is an average HR over 12 months), difference in restricted mean survival time ($\Delta \mathit{RMST}$), the difference in median survival time ($\Delta \mathit{MST}$), and  the difference in mean survival time ($\Delta \mu$). In the control group the mean survival time for all scenarios is 13.0 months and the cumulative incidence at 12 months is 39.7$\%$ for all scenarios.
    In the proportional hazards (PH), Early, and Late scenarios the control group has an exponential distribution; in the PH scenario the active group also follows an exponential distribution; in the Early and Late scenarios, the hazard function is log-linearly related to the hazards of the control group; in the null scenario both groups follow the same exponential distribution.}
\begin{tabular}{c c c c c c c} 
\toprule
   Scenario & $\WP(a,c)$ & $\WP_{\tau}(a,c)$ & $\HR$ & $\Delta \mathit{RMST}$ & $\Delta \mathit{MST}$ & $\Delta\mu$\\
        \midrule
\multirow{3}{*}{PH}& 0.61 & 0.58 & 0.65 & 1.19 & 4.85 & 6.99 \\ 
 & 0.56 & 0.55 & 0.80 & 0.65 &  2.25 & 3.25 \\ 
& 0.53 & 0.52 & 0.90 & 0.32 &  1.00 & 1.44 \\ 
\midrule
\multirow{3}{*}{Early} & 0.61 & 0.59 & 0.64 & 1.29 &  4.96 & 5.28 \\ 
 & 0.56 & 0.56 & 0.82 & 0.88 &  2.17 & 1.82 \\ 
 & 0.53 & 0.53 & 0.93 & 0.41 &  0.79 & 0.74 \\ 
 \midrule
\multirow{3}{*}{Late} & 0.61 & 0.57 & 0.63 & 0.91 &  5.81 & 10.22 \\ 
 & 0.56 & 0.53 & 0.82 & 0.37 & 1.77 & 6.87 \\ 
 & 0.53 & 0.51 & 0.91 & 0.17 & 0.68 & 3.61 \\ 
 \midrule
 Null & 0.50 & 0.50 & 1.00 & 0.00 & 0.00 & 0.00 \\ 
\bottomrule
\label{est_rel_table}
\end{tabular}
\end{table}

\subsection{Simulation results \label{sim_res}}



\begin{figure}[!tbp]
    \centering
   \includegraphics[width=0.4\textwidth]{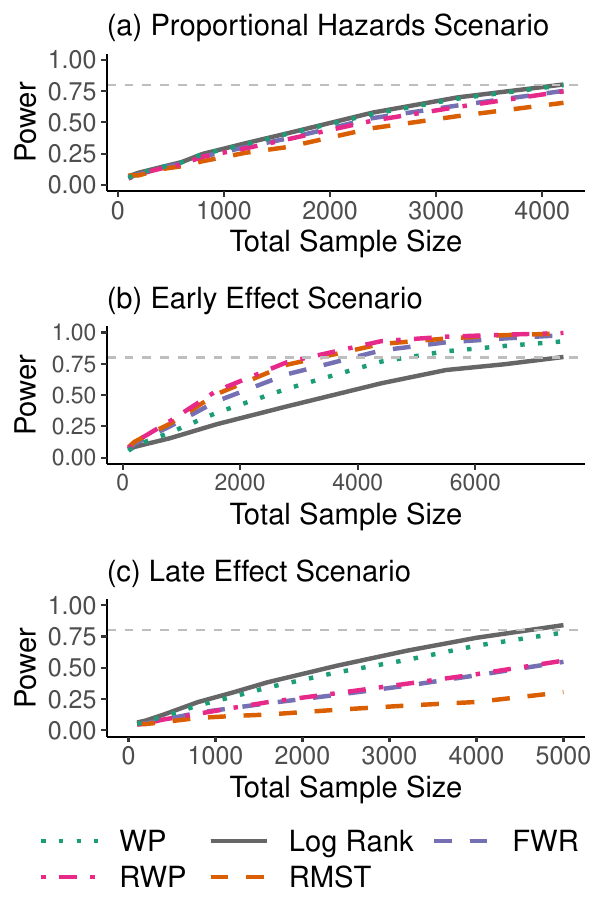}
   \caption{Power curves for all scenarios with a true $\WP(a,c)=0.526$ and a follow-up period of 12 months. Panel (a) is the proportional hazards scenario where both groups follow an exponential distribution with an event rate of 70\%. Panels (b) and (c) have control groups following an exponential distribution and treatment groups with log-linear hazards. In (b), the log-linear early scenario, there is an early treatment difference with an event rate of 71\%. In (c), the log-linear late scenario, there is a late treatment difference with an event rate of 69\%. The horizontal dashed lines indicate 80\% power.}
   \label{power_526_f12}
\end{figure}

The type I error rate for all methods is well controlled across scenarios, ranging from 0.044 to 0.058 (see Web Table 1, for a longer follow-up time see Web Table 2). Figure \ref{power_526_f12} depicts power curves with $\WP(a,c)=0.525$ for all scenarios with a minimum follow-up time of 12 months. In the PH scenario, panel (a), at $N=4200$, the power in decreasing order is 80.3\% for the log-rank test, 79.7\% for WP, 75.2\% for FWR, 74.7\% for RWP, and 65.6\% for RMST. In the early effects scenario, panel (b), at $N=2800$, the power in decreasing order is 76.4\% for RWP, 74.3\% for RMST, 66.7\% for FWR, 55.1\% for WP, and 41.0\% for the log-rank test. In the late effects scenario, panel (c), at $N=5000$, the power in decreasing order is 84.1\% for the log-rank test, 78.2\% for WP, 55.7\% for RWP, 54.8\% for FWR, and 30.5\% for RMST. These trends in power hold for all scenarios across sample sizes. In all scenarios, the FWR has lower or comparable power than the proposed RWP, and has a substantially lower power than the proposed WP in the late effect scenario. In all scenarios, either RWP or WP has higher power than RMST, and in the PH and late effect scenarios, both RWP and WP have higher power than RMST. The power curves for the two larger effect sizes, $\WP=0.556$ and 0.606 can be found in Web Figures 2-3, and the Power curves for a longer follow-up time can be found in Web Figures 4-6.


Table \ref{sim_res_tbl_12} depicts performance metrics for WP and RWP with $\tau=12$ and a follow-up time of 12 months. In all the scenarios we considered, the WP metric has a larger true effect size than the RWP measure. The relative bias for all scenarios and summary measures is small, $<0.01$, with similar magnitudes between measures. The WP metric is also expected to have more variability than the RWP metric as it will be affected by the tail behavior of the survival distributions. The RMSE for the WP summary measure is larger than the RWP by between 0.002 and 0.003 across all scenarios. The credible interval width is between 0.009-0.013 units smaller for RWP than WP.  The credible interval coverage is $<1\%$ different than the nominal level, 0.95, for all scenarios and summary measures.  The performance metrics for a longer follow-up time (Web Table 3) show a similar pattern as Table \ref{sim_res_tbl_12}.

\begin{table}[h]
\centering
    \caption{Performance metrics for win probability $\WP(a,c)$ and restricted win probability $\WP_{\tau}(a,c)$ with $\tau=12$ are shown for $N = 800$ and a minimum follow up of 12 months. The proportional hazards (PH) scenario has both groups following exponential distributions. The early effects (Early) and late effects (Late) scenarios have exponentially distributed control groups with active group hazards log-linearly related to the control group hazards, with early and late differences respectively. Relative bias (Bias) is the average difference between the estimated summary measure and the expected value. Root mean squared error (RMSE) is an accuracy measure. Credible interval coverage percentage (CI Cov \%) is the percentage of simulations where the 95\% CI includes the expected value of the summary measure. Credible interval width (CI Width) is the average length of the 95\% CI.}
    \addtolength{\tabcolsep}{-0.2em}
\begin{tabular}{l c c c c c c c c c c}
    \toprule
    &    \multicolumn{5}{c}{$\WP(a,c)$} & \multicolumn{5}{c}{$\WP_{\tau}(a,c)$}\\
    \cmidrule(lr){2-6}\cmidrule(lr){7-11}
     Scenario & Truth  & Bias$^\ddagger$ & RMSE$^\ddagger$& CI  & CI  & Truth&Bias$^\ddagger$  & RMSE$^\ddagger$ & CI  & CI \\ 
      &  &  & & Cov \% &  Width$^\ddagger$ &  &  & & Cov \% &  Width$^\ddagger$\\ 
      \midrule
PH & 0.526 & 0.02 & 2.18 & 94.6 & 8.33 & 0.522 & 0.02 & 1.94 & 94.6 & 7.38 \\ 
Early & 0.526 & -0.32 & 2.09 & 95.6 & 8.31 & 0.526 & 0.29 & 1.86 & 94.6 & 7.38 \\ 
Late & 0.526 & -0.44 & 2.20 & 94.3 & 8.35 & 0.513 & 0.48 & 1.95 & 94.8 & 7.40 \\ \midrule
PH & 0.556 & 0.14 & 2.17 & 95.0 & 8.38 & 0.545 & 0.13 & 1.89 & 95.0 & 7.30 \\ 
Early & 0.556 & -0.44 & 2.19 & 94.4 & 8.34 & 0.556 & -0.43 & 1.92 & 94.2 & 7.28 \\ 
Late & 0.556 & -0.56 & 2.13 & 95.2 & 8.42 & 0.529 & 0.99 & 1.89 & 94.8 & 7.33 \\ \midrule
PH & 0.606 & 0.21 & 2.15 & 94.8 & 8.35 & 0.583 & 0.24 & 1.82 & 94.5 & 7.10 \\ 
Early & 0.606 & 0.51 & 2.15 & 94.5 & 8.35 & 0.589 & -0.07 & 1.81 & 95.4 & 7.08 \\ 
Late & 0.606 & 0.79 & 2.23 & 94.2 & 8.38 & 0.569 & 0.60 & 1.87 & 94.2 & 7.15 \\ 
   \hline
\multicolumn{11}{l}{$^\ddagger$scaled by $10^{2}$}
      \label{sim_res_tbl_12}
\end{tabular}
\end{table}

\section{Data Applications}\label{real dat}

We analyzed a total of 304 datasets reconstructed from 153 oncology trial publications with time-to-event endpoints in the \texttt{KMdata} R package by using the Guyot algorithm \citep{guyot_enhanced_2012}. 
The following types of cancer were studied in these trials: breast (110 publications), lung (87), colorectal (68), prostate (37), and lung/colorectal (2). The majority (88.5\%) of these datasets are from superiority trials. The median reported sample size of datasets is 608 (range: 78 to 8,381) with a median follow-up time of 48 months (range: 0.4 to 240 months). The two most common survival outcomes were overall survival (46.4\%) and progression free survival (34.9\%). The median of the estimated survival at the last follow-up time was 19.7\%, with a range of 0\% to 95.5\%. 
The Grambsch-Therneau test indicated the PH assumption holds in 247 (81.25\%) datasets \citep{grambsch_proportional_1994}. RWP rather than WP is compared with various existing methods as the measure allows for ties in the instance that the survival curves are incomplete.

\begin{figure}
    \centering
    \includegraphics[width=0.7\textwidth]{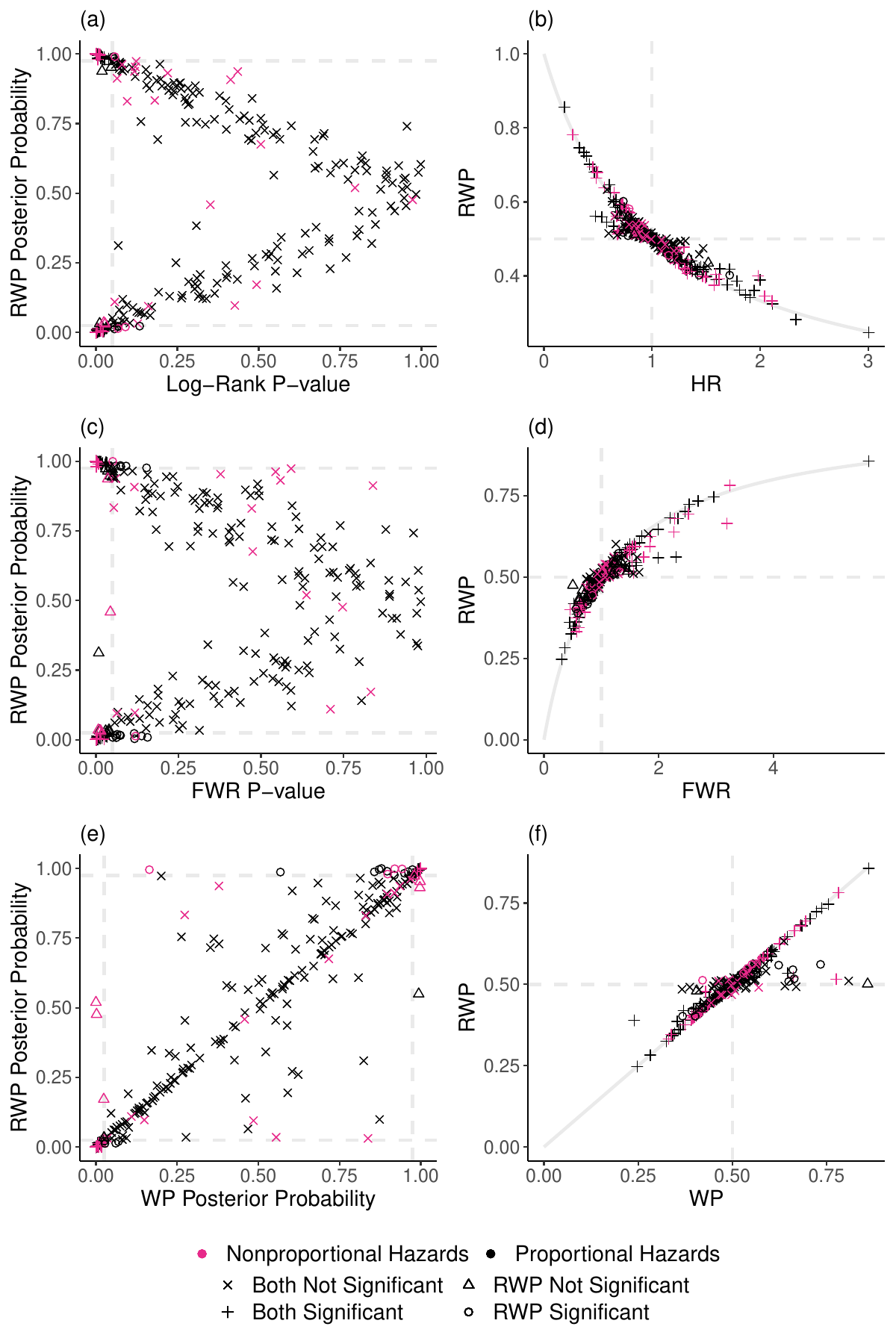}
 \caption{Significance and treatment effect estimates for 
 the 304 reconstructed datasets between the proposed restricted win probability (RWP) method and the log-rank test (the upper panels), the frequentist win ratio (FWR) test (the middle panels), and the proposed WP method (the lower panels). Panel (a) depicts the log-rank p-values and $\RWP$ posterior probabilities (both are used to test for treatment efficacy), where the dashed lines indicate the thresholds for significance; (b) depicts the treatment effect estimates of hazard ratio (HR) and $\RWP$, where the dashed lines indicate the null effect for $\HR$ and $\RWP$, and the grey line indicates the relationship, $\WP=1/(1+\HR)$, under proportional hazards; (c) depicts the FWR method's p-values and $\RWP$ posterior probabilities (both are used to test for treatment efficacy), where the dashed lines indicate the thresholds for significance; (d) depicts the treatment effect estimates of $\FWR$ and $\RWP$, where the dashed lines indicate the null effect for $\FWR$ and $\RWP$, and the grey line indicates the relationship, $\FWR=\WP/(1-\WP)$; (e) depicts the $\WP$ and $\RWP$ posterior probabilities where the dashed lines indicate thresholds for significance; (f) depicts the treatment effect estimates of $\WP$ and $\RWP$ where the dashed lines indicate the null effect for $\WP$ and $\RWP$, and the grey line indicates the relationship $\RWP = \WP$.}
\label{fig3_kmdata_comp}
\end{figure}

Reguarding a comparison between the log-rank test and RWP, the treatment effect measure for log-rank test, the HR, was estimated by using a Cox model with Breslow's method for ties. For the RWP, $\tau$ was set to the last observed event time.
Figure \ref{fig3_kmdata_comp}a shows that the proposed RWP method and the log-rank test agree on group differences in 286 (94.1\%) of the datasets. Among the 18 datasets with discordant conclusions, 7 had a significant log-rank test but a non-significant result using RWP, and 11 had a significant result using RWP but a non-significant log-rank test. Moreover, 5 (27.8\%) exhibited non-PH (see Web Figure 7 for their Kaplan-Meier curves). Figure \ref{fig3_kmdata_comp}b shows that $\RWP$ and $\HR$ have an inverse relationship. In fact, under proportional hazards, the  unrestricted WP is inversely related to the HR as follows, $\mathit{WP}=\frac{1}{\mathit{HR}+1}$ \citep{gonen_concordance_2005}.
This relationship may aid in calibrating hypothesized effect sizes for the proposed measures and HRs commonly reported in the literature.

In relation to the RWP and FWR, Figure \ref{fig3_kmdata_comp}c shows that the proposed method, RWP, and the FWR agree on the group difference in 270 (88.8\%) of the datasets, whereas in 17 datasets the RWP method concludes a significant difference between groups while the FWR does not, and vice versa for the other 17 datasets. In Figure \ref{fig3_kmdata_comp}d shows a positive relationship between $\FWR$ and $\RWP$. We note that by definition, $\FWR = \WP/(1-\WP)$. The Kaplan-Meier curves of the 34 datasets with disagreements can be found in Web Figure 8.

In regards to the WP and RWP, Figure \ref{fig3_kmdata_comp}e shows that the proposed methods, RWP and WP, agree on the group difference in 279 (91.8\%) of the datasets, whereas in 17 datasets the RWP method concludes a significant difference between groups while the WP does not, and vice versa for the other 8 datasets. In Figure \ref{fig3_kmdata_comp}f, the relationship between the $\WP$ and $\RWP$ values appears mostly linear. The Kaplan-Meier curves of the 25 datasets with disagreements can be found in Web Figure 9. 

\captionsetup[subfigure]{position=top, format=hang, justification=centering}

\captionsetup[subfigure]{position=top, format=hang, justification=centering}
\begin{figure}
    \centering
    \includegraphics[width=0.8\textwidth]{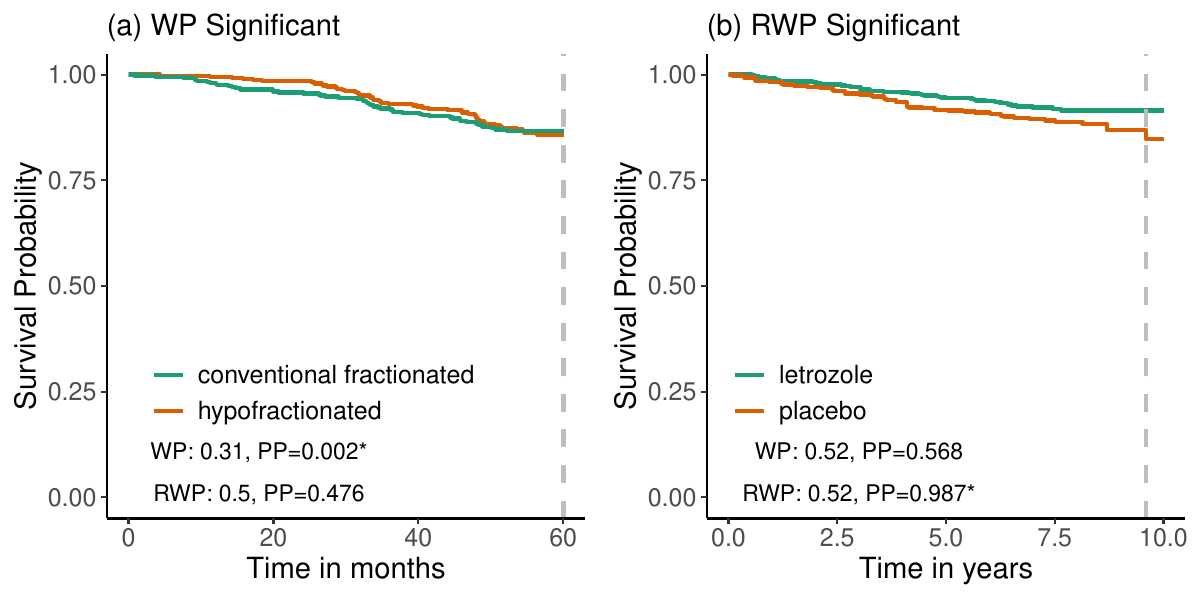}
 \caption{Examples of differences in significance between the proposed win probability (WP) and restricted win probability (RWP)  methods, with large differences in posterior probability (PP). Panel (a) shows Kaplan-Meier curves of overall survival (a secondary endpoint) from the HPYRO Trial,  \cite{incrocci_hypofractionated_2016} Figure 2B,
 which compared hypo-fractional radiotherapy and conventionally fractioned radiotherapy in 804 patients with localized prostate cancer; 
 the original analysis of the trial did not conclude a significant difference between groups. Panel (b) shows Kaplan-Meier curves of disease-free survival (the primary endpoint) from the MA.17R Trial,  \cite{goss_extending_2016} Figure 1A,
 which compared letrozole and placebo in 1918 breast cancer patients; 
 the original analysis did conclude a significant difference. The dashed vertical line indicates the $\tau$ used for RWP. $^{*}$ indicates statistical significance.}
\label{kmdata_rwp_wp_kmcurves}
\end{figure}

Figure \ref{kmdata_rwp_wp_kmcurves} shows Kaplan-Meier curves from two case studies where WP and RWP disagree on significance and have some of the largest differences in posterior probability. In Figure \ref{kmdata_rwp_wp_kmcurves}a, WP shows a significant difference between groups while RWP does not. This large difference between summary measures may be due to tail behavior affecting the WP measure which may be largely influenced by the two survival curves crossing. In Figure \ref{kmdata_rwp_wp_kmcurves}b, RWP concludes a significant difference while WP does not, even though both have the same treatment effect estimate; in this case the large difference in posterior probability is likely due to the larger variability of the WP measure. In both cases, RWP agrees in significance with the original trials analysis.

\section{Discussion} \label{discussion}

The proposed restricted and unrestricted WP estimands with Bayesian estimation offer a new way to compare survival curves in the context of clinical trials with univariate survival outcomes within the estimand framework. 
In the simulation study, we showed that the RWP approach had similar power to RMST difference and higher power than the log-rank test when there was an early treatment effect that diminishes over time. In scenarios with proportional hazards and late treatment effects, the WP approach had similar power to the log-rank test. In these PH scenarios, the RWP approach had higher power than RMST difference. The RWP approach performed most poorly when there is a delayed treatment effects, which may reflect its similarity to Wilcoxon tests \citep{martinez_pretest_2010}. In all scenarios, the WP approach has a larger (or similar) treatment effect size when compared to the RWP approach, but has larger variability. In a comparison of different approaches for censored data for generalized pairwise comparisons, all methods lead underestimated the net benefit under administrative censoring \cite{deltuvaite-thomas_generalized_2023}. Our methods show similar bias for the two different levels of administrative censoring.


We considered a Bayesian model that assumed a flexible two-parameter Weibull distribution for the time-to-event outcome in each randomization group. 
This approach affords greater flexibility regarding the nature of the treatment effect than many popular parametric and semi-parametric time-to-event regression approaches (e.g., Cox PH and AFT models). 
The proposed estimands may be estimated with any Bayesian model suitable for time-to-event outcomes, including fully non-parametric approaches \citep{damien_bayesian_2002}. More flexible parametric models include a generalized gamma model or generalized F \citep{cox_parametric_2007, cox_generalized_2008}. 
We estimated the proposed WP estimands with the Weibull model as it is easy to implement in JAGS and Stan and may fit the observed data well in contexts where the hazard function is expected to be monotone. 

We only discussed using the WP metric in the context of time-to-event outcomes with right censoring. Due to our adoption of Bayesian methods, our proposed method can be easily extended to allow for 
other forms of censoring and truncation such as interval censored survival data or left-truncated and right-censored data that are frequently encountered in clinical research \citep{damle_ig_1999, hamblin_unmutated_1999, rodrigues_use_2018}. 
These alternative types of censored data will only impact Bayesian model fitting, not the calculation of the proposed summary measures conditional on the fitted model. 
Another useful extension would be to allow for covariate adjustment or stratification in treatment comparisons \citep{gasparyan_adjusted_2021}. 

As with RMST, for the RWP the choice of the restriction time $\tau$ will impact the estimand and estimates, and thus it should be pre-specified. When everyone has the same administrative follow-up time, one can use that for $\tau$. However, when participants have varying lengths of administrative follow-up, $\tau$ should be less than or equal to the longest observed follow-up time. Similar to options suggested for RMST, $\tau$ could be selected based on clinical relevance or data maturity, and one could report sensitivity analyses across a range of values for $\tau$ \citep{royston_restricted_2013}.

\section*{Acknowledgement}
Luo's research is partially supported by the National Cancer Institute's Cancer Center Support Grant (P30 CA77598) to the University of Minnesota Masonic Cancer Center and the University of Minnesota’s NIH Clinical and Translational Science Award by the National Institute of Health (UM1TR004405).
\section*{Conflicts of Interest}
None.

\bibliography{wp_bib_10_6_2023}

\end{document}


\title{Supplementary Material for\\
Restricted Win Probability with Bayesian Estimation for Implementing the Estimand Framework in Clinical Trials With a Time-to-Event Outcome}
\author
{by Michelle Leeberg,
Xianghua Luo,
Thomas A. Murray}
\date{}
\maketitle

\subsection*{Web Appendix A}
\subsubsection*{A.1 Details for Deriving the Restricted Win Probability}
\begin{align}
\tag{A1}
\begin{aligned}
\WP_{\tau}(a,c)&=\E[r_{\tau}(Y_{a},Y_{c})]\\
& = \Pr(Y_{a}>Y_{c} \text{ and } Y_{c} \leq \tau) + 0.5\Pr(\min(Y_{a},Y_{c}) > \tau)\\
& = \int_{0}^{\tau}\{1-F_{a}(t)\}f_{c}(t)dt + 0.5\{1-F_{a}(\tau)\}\{1-F_{c}(\tau)\}\\
& = \int_{0}^{\tau}S_{a}(t)f_{c}(t)dt + 0.5 S_{a}(\tau) S_{c}(\tau),
\end{aligned}
\end{align}    
since
\begin{align}
\tag{A2}
\begin{aligned}
    \Pr(Y_{a}>Y_{c} \text{ and } Y_{c} \leq \tau) & = \int_{0}^{\tau}\int_{y_{c}}^{\infty}f_{a,c}(y_{a},y_{c})dy_{a}dy_{c}\\
 & = \int_{0}^{\tau}\int_{y_{c}}^{\infty}f_{a}(y_{a})f_{c}(y_{c})dy_{a}dy_{c}\\
  & = \int_{0}^{\tau}\Big\{\int_{y_{c}}^{\infty}f_{a}(y_{a})dy_{a}\Big\}f_{c}(y_{c})dy_{c}\\
    & = \int_{0}^{\tau}\{1-F_{a}(y_{c})\}f_{c}(y_{c})dy_{c}\\
        & = \int_{0}^{\tau}\{1-F_{a}(t)\}f_{c}(t)dt,
\end{aligned}
\end{align}
and
\begin{align}
\tag{A3}
\begin{aligned}
 \Pr(\min(Y_a,Y_{c})>\tau)&=\Pr(Y_a>\tau,Y_{c}>\tau) \\
 &= \Pr(Y_a>\tau)\Pr( Y_{c}>\tau)\\
 &=\{1-F_{a}(\tau)\}\{1-F_{c}(\tau)\}.  
\end{aligned}
\end{align}

\subsection*{Web Appendix B}
\subsubsection*{B.1 Simulation Scenarios}

All scenarios have a control group following an exponential distribution with a median survival time of 9 months. Scenario 1, PH scenario, contains data simulated with exponential PH models. In the control group, $h_{c}(t)=\lambda$, while in the active treatment group, $h_{a}(t)=\HR\cdot\lambda$; $\HR = 0.65, 0.8, 0.9$. 

Scenarios 2 and 3 contain data simulated with a log-linear hazard in the active group. Control data are simulated using an exponential distribution with hazard, $h_{c}(t)=\lambda$. Scenario 2, early effects scenario, has log-linear hazards in the treatment group with early treatment differences. In this scenario the HR starts at 0.6, with the treatment hazard increasing log-linearly over time until the HR reaches 1. This gives the hazard in the treatment group the form shown as follows, with $\psi=0.00042, 0.03504, 0.1010$: 

\begin{equation}
\tag{A4}
   h_{a}(t, \psi) = \left\{
     \begin{array}{lr}
       0.6\lambda \exp(\psi\cdot t) & ,\ t\leq -\psi^{-1}\log (0.6), \\ 
      \lambda & ,\ t> -\psi^{-1}\log (0.6). \\
     \end{array}
   \right.
\end{equation} 

Scenario 3, late effects scenario, has a log-linear hazard in the treatment group with late treatment differences. Here, the HR starts at 1, with the treatment hazard decreasing log-linearly over time, until the HR reaches 0.5. The hazard in the treatment group is shown as follows, with $\zeta=-1.1246, -0.0962, -0.0344$:

\begin{equation}
\tag{A5}
   h_{a}(t, \zeta) = \left\{
     \begin{array}{lr}
       \lambda \exp(\zeta\cdot t) & ,\ t\leq \zeta^{-1}\log (0.5), \\ 
      0.5\lambda & ,\ t> \zeta^{-1}\log (0.5) .\\
     \end{array}
   \right.
\end{equation}

For each scenario above, null scenarios where both groups follow the same distribution will be used to find two-sided type I error rates (see Web Tables 1 and 2). As all scenarios have a control group with an exponential distribution, a single scenario with both groups following an exponential distribution with a median survival time of 9 months will be used for the control scenario. Further, for each scenario and $\psi$ or $\zeta$ value, a null scenario is created with both groups following the same distribution of the active treatment group from the corresponding non-null scenarios ($\WP(a,c)=0.526,0.556,0.606$) presented in Table 2 of the manuscript.

Each scenario had 10 sample sizes ranging from 100 to their maximum sample sizes. The maximum sample sizes for the PH, early differences, and late differences scenarios were 4000, 10500, and 1600, respectively. For each scenario, 2000 data sets were simulated. 

\begin{figure}[!htbp]
\centering
\includegraphics[width=0.9\textwidth]{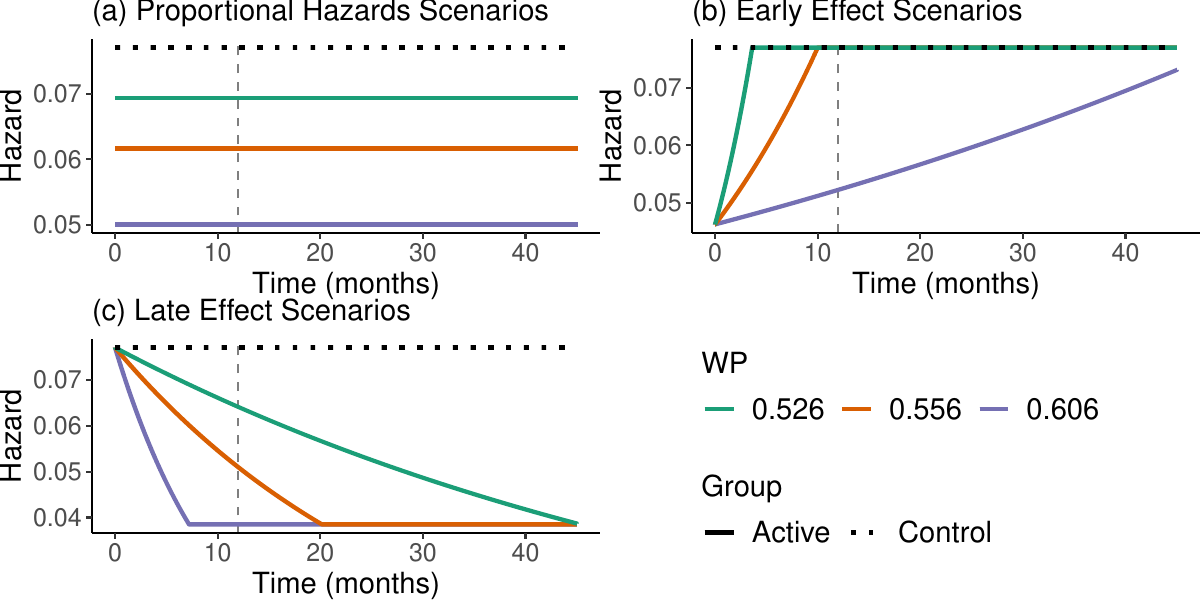}
   \caption*{Web Figure 1. Hazard curves for all scenarios. In panel (a), treatment groups follow exponential distributions with proportional hazards to the control group. In panels (b) and (c), treatment groups have log-linear hazard functions; with (b) having early differences in hazard and (c) having late differences in hazard. Within each plot, three levels of treatment effect as measured by the win probability (WP) are presented, while the vertical dashed line indicates the follow-up time of 12 months, $\tau=12$ for studying the restricted WP.}
\end{figure}

\subsubsection*{B.2 Additional Simulation Information}

All survival times are simulated by finding  $F^{-1}(u)$ with $u \sim U(0,1)$ \cite{bender_generating_2005, crowther_simulating_2013}. For simulations, all parameters are initiated at the true value. For the JAGS model, 3 MCMC chains each with 5000 iterations was used.
For the RMST estimation, 
$\tau$, the time to estimate mean survival time to, is set to the minimum of follow-up time and the minimum of the maximum event time or censoring time by group. 
Additional simulation information can be found in 
\href{https://github.com/msonnenb110}{https://github.com/msonnenb110}.

\subsubsection*{B.3 MCMC Estimation}
For each model, we pull $x$ draws from $c$ MCMC chains, leading to $B=x\cdot c$ total draws. Using these posterior draws, we can obtain the distributions of $Y_{a}$ and $Y_{c}$ given a particular dram for the model parameter. For each draw, we can then obtain the model-based win probability estimate by plugging posterior estimates of the survival and probability density distributions into manuscript Equations (3) and (5), e.g., $\WP(a,c;\bm{\theta}_j)$ for $j=1,...,B$. We then estimate the posterior probability that the particular win probability summary measure is greater than 0.5 as follows:
\begin{equation}
\tag{A6}
    \Pr(\WP(a,c;\boldsymbol{\theta}) > 0.5 | \textrm{Data})=\frac{1}{B}\sum_{j=1}^{B}\I(\WP(a,c;\bm{\theta}_j)>0.5).
\end{equation}

\newpage
\subsubsection*{B.4 Additional Simulation Results}
\subsubsection*{12 Month Follow Up}

\begin{table}[!htbp]
\centering
    \caption*{Web Table 1. Two-sided type I error rate across different null scenarios with a minimum follow-up time of 12 months, and $\tau=12$. In all scenarios, the control group follows the active group's distribution. In the ``Control Exp'' row, both groups follow the control group's exponential distribution. For all rows other than the ``Control Exp'' row, the two groups' distribution is 
    based on the active group's distribution in the corresponding scenarios of Table 2 in the manuscript. 
    The following metrics are presented: win probability (WP), restricted win probability (RWP), log-rank test (LR), restricted mean survival time (RMST), and frequentist win ratio (FWR). The WP Effect Size indicates the active group used for the null scenario. }
\begin{tabular}{ c c c c c c c}
   \toprule
 Scenario &  $\Pr(Y_{a}>Y_{c})$ & \multicolumn{5}{c}{Type I Error Rate}\\ \cmidrule(lr){3-7}
  &  Effect Size &WP & RWP & LR & RMST & FWR \\ 
      \midrule
Control Exp &- & 0.058 & 0.054 & 0.060 & 0.049 & 0.054 \\ 
\hline
\multirow{3}{*}{PH} & 0.526 & 0.055 & 0.055 & 0.053 & 0.059 & 0.061 \\ 
 & 0.556 & 0.051 & 0.050 & 0.048 & 0.048 & 0.050 \\ 
 & 0.606 & 0.057 & 0.051 & 0.056 & 0.053 & 0.052 \\ 
  \hline
\multirow{3}{*}{Early effect} & 0.526 & 0.049 & 0.049 & 0.049 & 0.046 & 0.044 \\ 
 & 0.556 & 0.051 & 0.053 & 0.053 & 0.048 & 0.045 \\ 
 & 0.606 & 0.047 & 0.056 & 0.048 & 0.050 & 0.046 \\ 
  \hline
\multirow{3}{*}{Late effect} & 0.526 & 0.052 & 0.056 & 0.051 & 0.050 & 0.053 \\ 
 & 0.556 & 0.049 & 0.044 & 0.044 & 0.044 & 0.045 \\ 
 & 0.606 & 0.053 & 0.053 & 0.052 & 0.045 & 0.044 \\ 
   \bottomrule
\end{tabular}
\label{typeI_f12}
\end{table}

\begin{figure}[!htbp]
    \centering
   \includegraphics[width=0.5\textwidth]{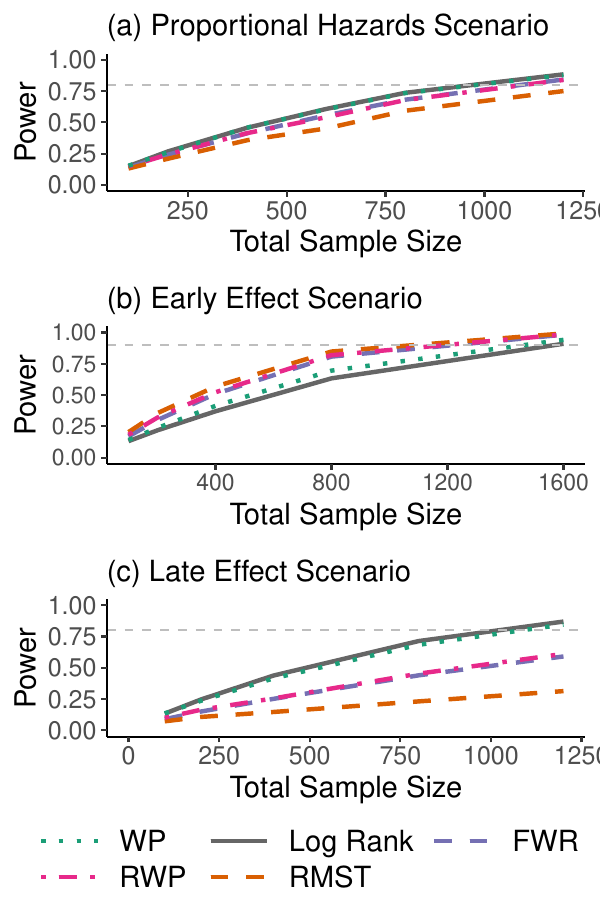}
   \caption*{Web Figure 2. Power curves for all scenarios with a true $\WP(a,c)=0.556$, with a minimum follow up of 12 months ($\tau$=12 for RWP). Panel (a) is the proportional hazards scenario where both groups follow an exponential distribution with an event rate of 67\%. Panels (b) and (c) have control groups following an exponential distribution and treatment groups with log-linear hazards. In (b), the early differences scenario, there is an early treatment difference with an event rate of 69\%. In (c), the late differences scenario, there is a late treatment difference with an event rate of 66\%.}
   \label{pw_556_f12}
\end{figure}

\begin{figure}[!htbp]
\centering
   \includegraphics[width=0.5\textwidth]{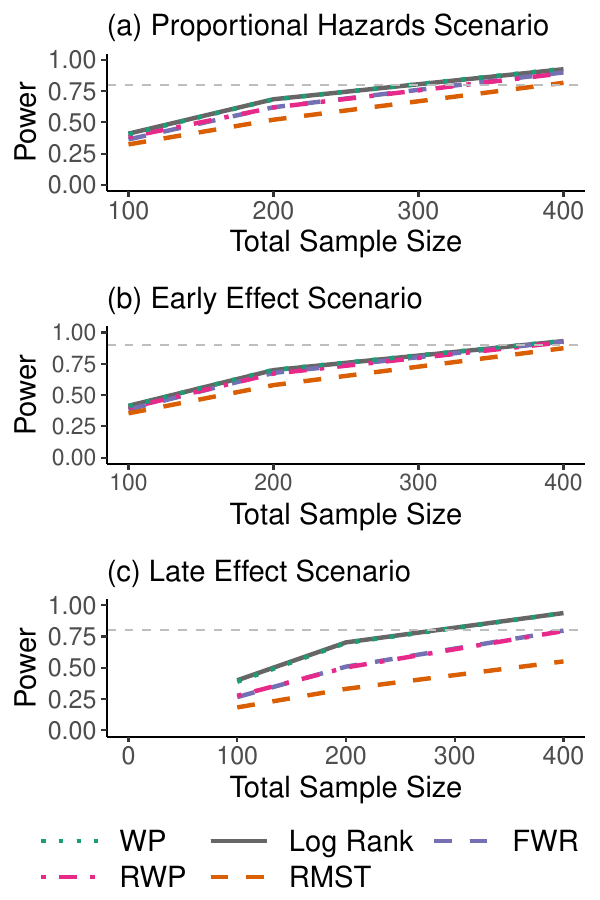}
   \caption*{Web Figure 3. Power curves for all scenarios with a true $\WP(a,c)=0.606$, with a minimum follow up of 12 months ($\tau$=12 for RWP). Panel (a) is the proportional hazards scenario where both groups follow an exponential distribution with an event rate of 64\%. Panels (b) and (c) have control groups following an exponential distribution and treatment groups with log-linear hazards. In (b), the early differences scenario, there is an early treatment difference with an event rate of 64\%. In (c), the late differences scenario, there is a late treatment difference with an event rate of 62\%.}
   \label{pw_606_f12}
\end{figure}

\newpage
\subsubsection*{36 Month Follow Up}

\begin{table}[!htbp]
\centering
    \caption*{Web Table 2. Two-sided type I error rate across different null scenarios with a minimum follow-up time of 36 months, and $\tau=36$. In all scenarios, the control group follows the active group's distribution. In the ``Control Exp'' row, both groups follow the control group's exponential distribution. For all rows other than the ``Control Exp'' row, the two groups' distribution is 
    based on the active group's distribution in the corresponding scenarios of Table 2 in the manuscript.
    The following metrics are presented: win probability (WP), restricted win probability (RWP), log-rank test (LR), restricted mean survival time (RMST), and frequentist win ratio (FWR). The WP Effect Size indicates the active group used for the null scenario. }
\begin{tabular}{ c c c c c c c}
   \toprule
 Scenario &  $\Pr(Y_{a}>Y_{c})$ & \multicolumn{5}{c}{Type I Error Rate}\\ \cmidrule(lr){3-7}
  &  Effect Size &WP & RWP & LR & RMST & FWR \\ 
      \midrule
Control Exp &- & 0.048 & 0.048 & 0.050 & 0.047 & 0.047 \\ 
\hline
\multirow{3}{*}{PH} & 0.526 & 0.051 & 0.051 & 0.049 & 0.051 & 0.052 \\ 
 & 0.556 & 0.043 & 0.043 & 0.043 & 0.045 & 0.044 \\ 
 & 0.606 & 0.053 & 0.053 & 0.054 & 0.055 & 0.053 \\ 
\hline
\multirow{3}{*}{Early} & 0.526 & 0.058 & 0.058 & 0.060 & 0.061 & 0.057 \\ 
 & 0.556 & 0.048 & 0.048 & 0.047 & 0.048 & 0.043 \\ 
 & 0.606 & 0.050 & 0.050 & 0.047 & 0.048 & 0.047 \\ 
\hline
\multirow{3}{*}{Late} & 0.526 & 0.041 & 0.041 & 0.052 & 0.050 & 0.041 \\ 
 & 0.556 & 0.047 & 0.047 & 0.051 & 0.049 & 0.050 \\ 
 & 0.606 & 0.052 & 0.052 & 0.047 & 0.052 & 0.052 \\ 
   \bottomrule
\end{tabular}
\label{typeI_f36}
\end{table}

\begin{figure}[!htbp]
    \centering
   \includegraphics[width=0.5\textwidth]{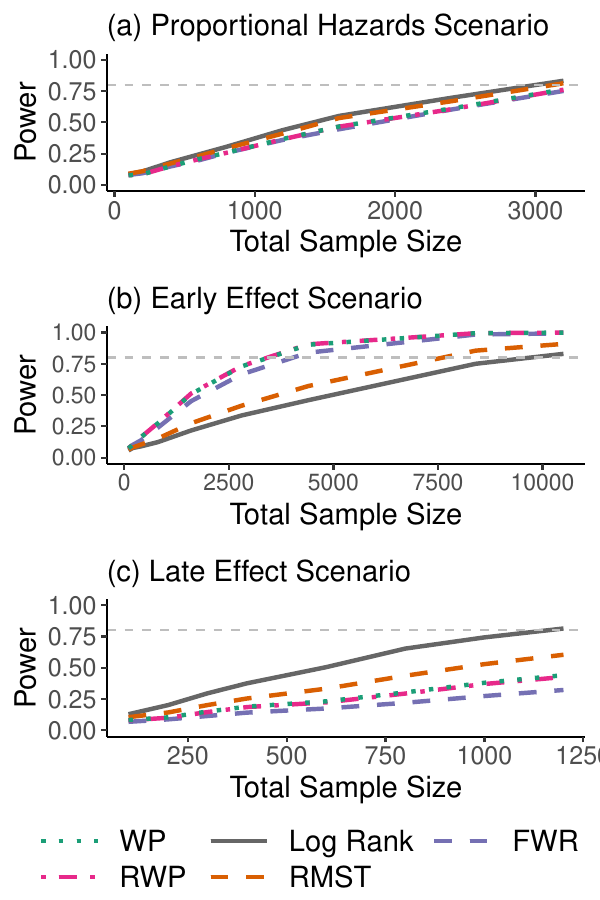}
   \caption*{Web Figure 4. Power curves for all scenarios with a true $\WP(a,c)=0.526$, with a minimum follow up of 36 months ($\tau$=36 for RWP). Panel (a) is the proportional hazards scenario where both groups follow an exponential distribution with an event rate of 95\%. Panels (b) and (c) have control groups following an exponential distribution and treatment groups with log-linear hazards. In (b), the early differences scenario, there is an early treatment difference with an event rate of 95\%. In (c), the late differences scenario, there is a late treatment difference with an event rate of 93\%.}
   \label{pw_526_f36}
\end{figure}

\begin{figure}[!htbp]
    \centering
   \includegraphics[width=0.5\textwidth]{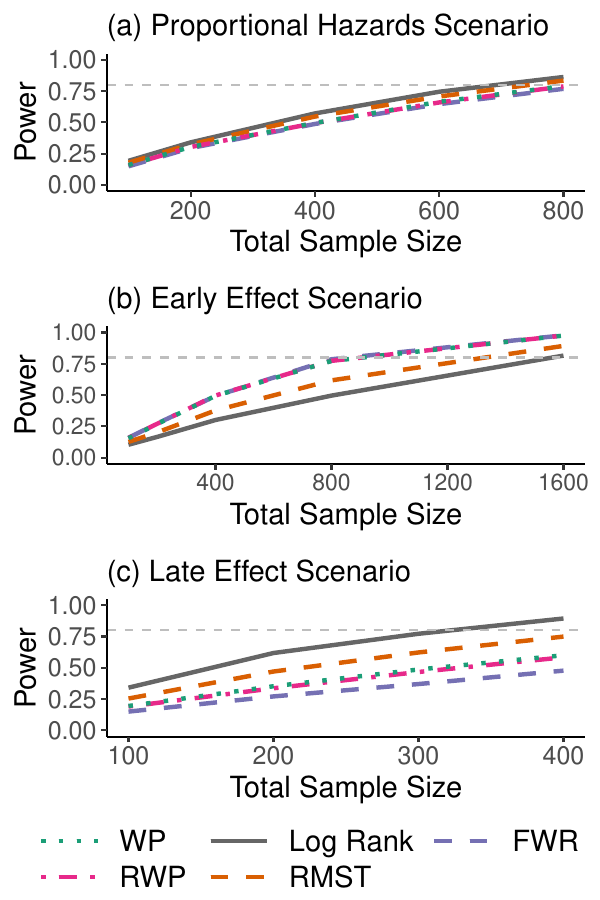}
   \caption*{Web Figure 5. Power curves for all scenarios with a true $\WP(a,c)=0.556$, with a minimum follow up of 36 months ($\tau$=36 for RWP). Panel (a) is the proportional hazards scenario where both groups follow an exponential distribution with an event rate of 94\%. Panels (b) and (c) have control groups following an exponential distribution and treatment groups with log-linear hazards. In (b), the early differences scenario, there is an early treatment difference with an event rate of 95\%. In (c),  the late differences scenario, there is a late treatment difference with an event rate of 90\%.}
   \label{pw_556_f36}
\end{figure}

\begin{figure}[!htbp]
\centering
   \includegraphics[width=0.5\textwidth]{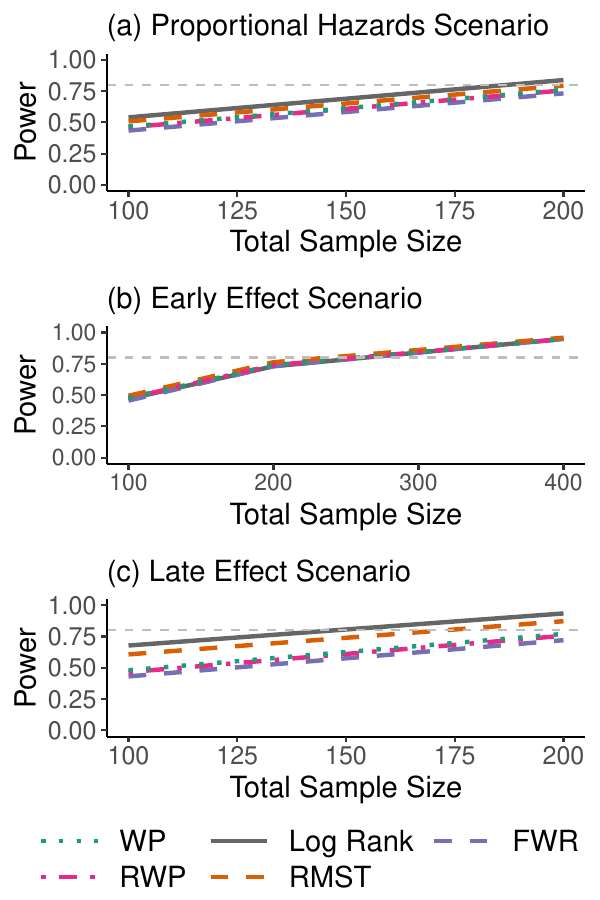}
   \caption*{Web Figure 6. Power curves for all scenarios with a true $\WP(a,c)=0.606$, with a minimum follow up of 36 months ($\tau$=36 for RWP). Panel (a) is the proportional hazards scenario where both groups follow an exponential distribution with an event rate of 91\%. Panels (b) and (c) have control groups following an exponential distribution and treatment groups with log-linear hazards. In (b), the early differences scenario, there is an early treatment difference with an event rate of 93\%. In (c), the late differences scenario, there is a late treatment difference with an event rate of 88\%.}
   \label{pw_606_f36}
\end{figure}

\begin{table}[!htbp]
\centering
    \caption*{Web Table 3. Performance metrics for win probability $\WP(a,c)$ and restricted win probability $\WP_{\tau}(a,c)$ with $\tau=36$ are shown for $N = 800$ and a minimum follow up of 36 months. The proportional hazards (PH) scenario has both groups following exponential distributions. The early effects (Early) and late effects (Late) scenarios have exponentially distributed control groups with active group hazards log-linearly related to the control group hazards, with early and late differences, respectively. Relative bias (Bias) was the average difference between the estimated summary measure and the expected value. Root mean squared error (RMSE) was an accuracy measure. Confidence/Credible interval coverage percentage (CI Cov \%), was the percentage of simulations where the 95\% CI includes the expected value of the summary measure. Credible interval width (CI Width) was the average length of the 95\% CI.}
    \addtolength{\tabcolsep}{-0.2em}
\begin{tabular}{l c c c c c c c c c c}
    \toprule
    &    \multicolumn{5}{c}{$\WP(a,c)$} & \multicolumn{5}{c}{$\WP_{\tau}(a,c)$}\\
    \cmidrule(lr){2-6}\cmidrule(lr){7-11}
     Scenario & Truth  & Bias$^\ddagger$ & RMSE$^\ddagger$& CI  & CI  & Truth&Bias$^\ddagger$  & RMSE$^\ddagger$ & CI  & CI \\ 
      &  &  & & Cov \% &  Width$^\ddagger$ &  &  & & Cov \% &  Width$^\ddagger$\\ 
      \midrule
PH & 0.526 & -0.03 & 2.04 & 94.5 & 7.85 & 0.526 & -0.03 & 2.04 & 94.6 & 7.84 \\ 
Early & 0.526 & 0.30 & 2.01 & 95.0 & 7.85 & 0.526 & 0.32 & 2.01 & 95.0 & 7.84 \\ 
Late & 0.526 & 0.76 & 2.00 & 94.4 & 7.86 & 0.525 & 0.86 & 2.01 & 94.4 & 7.84 \\ \midrule
PH & 0.556 & -0.02 & 1.98 & 94.7 & 7.81 & 0.555 & -0.02 & 1.98 & 94.7 & 7.79 \\ 
Early & 0.556 & -0.24 & 1.99 & 95.2 & 7.80 & 0.556 & -0.22 & 1.98 & 95.2 & 7.79 \\ 
Late & 0.556 & 1.40 & 2.13 & 93.6 & 7.82 & 0.554 & 1.44 & 2.13 & 93.7 & 7.80 \\ \midrule
PH & 0.606 & 0.13 & 1.89 & 96.2 & 7.65 & 0.605 & 0.13 & 1.88 & 96.3 & 7.64 \\ 
Early & 0.606 & -0.39 & 2.00 & 94.5 & 7.66 & 0.606 & -0.45 & 2.00 & 94.5 & 7.65 \\ 
Late & 0.606 & 0.15 & 1.92 & 95.6 & 7.68 & 0.604 & 0.09 & 1.92 & 95.5 & 7.66 \\ 
   \hline
      \label{sim_res_tbl_12}
\end{tabular}
 \begin{tablenotes}
    \item[3] $^\ddagger$scaled by $10^{2}$
  \end{tablenotes}
\end{table}


\captionsetup[subfigure]{position=top, format=hang, justification=centering}
\begin{figure}[!htpb]
    \centering
    \includegraphics[width=0.9\textwidth]{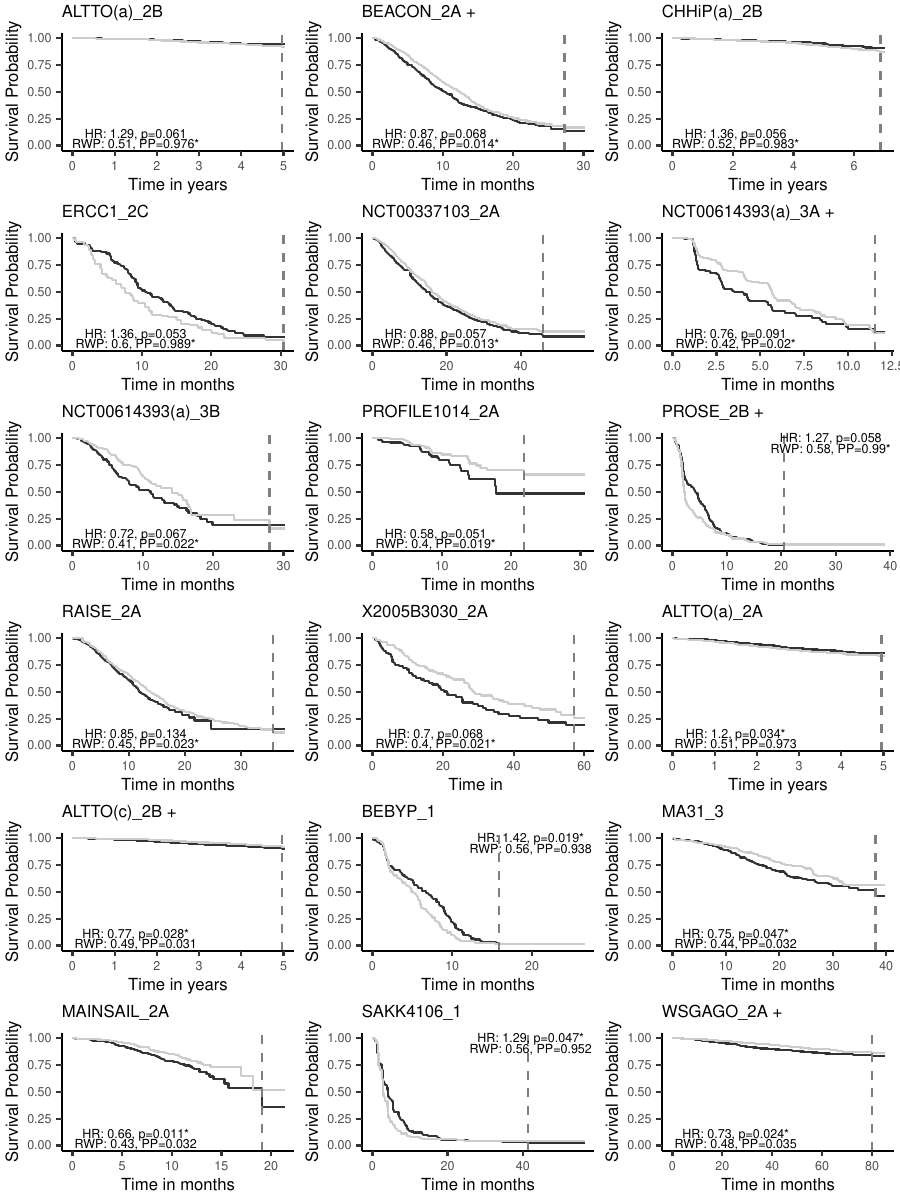}
 \caption*{Web Figure 7. Kaplan-Meier curves from the reconstructed datasets where the proposed restricted win probability (RWP) and log-rank test, with a hazard ratio as the summary measure, disagree on significance between treatment groups. $^{*}$ indicates significance; + indicates non-proportional hazards; dashed line indicates $\tau$ used for RWP.}
 \label{rwp_lr_dis}
\end{figure}

\begin{figure}[!htpb]
    \centering
    \includegraphics[width=0.85\textwidth]{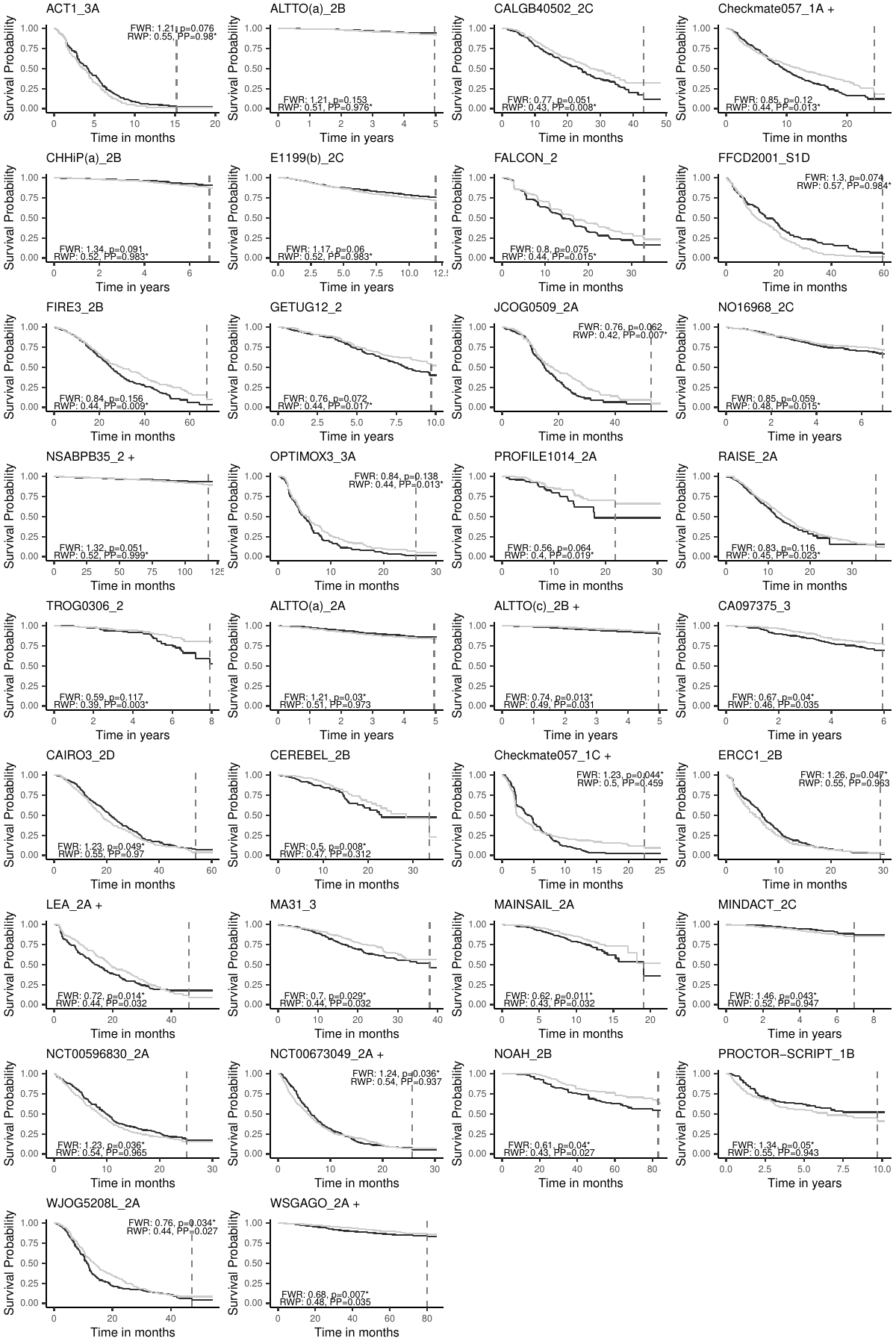}
 \caption*{Web Figure 8. Kaplan-Meier curves from the reconstructed datasets where the proposed restricted win probability (RWP) and frequentist win ratio (FWR) disagree on significance between treatment groups. $^{*}$ indicates significance; + indicates non-proportional hazards; dashed line indicates $\tau$ used for RWP.}
 \label{rwp_wr_dis}
\end{figure}

\begin{figure}[!htpb]
    \centering
    \includegraphics[width=\textwidth]{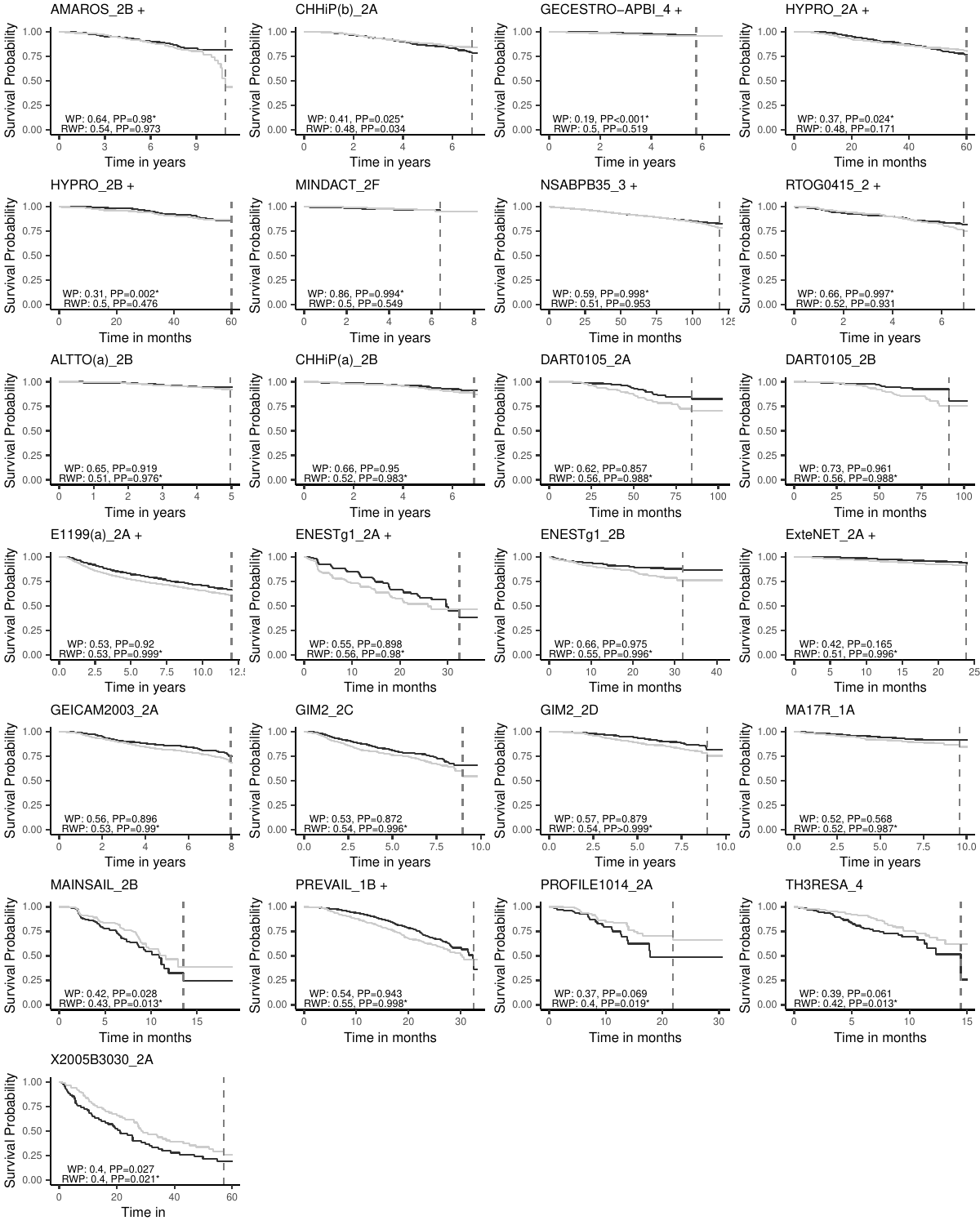}
 \caption*{Web Figure 9. Kaplan-Meier curves from the reconstructed datasets where the proposed restricted win probability (RWP) and win probability (WP) disagree on significance between treatment groups. $^{*}$ indicates significance; + indicates non-proportional hazards; dashed line indicates $\tau$ used for RWP.}
 \label{rwp_wp_dis}
\end{figure}

\newpage
\bibliography{wp_bib_10_6_2023}